\documentclass[12pt,preprint]{aastex}
\usepackage{longtable}
\usepackage{amsmath}
\usepackage{graphicx}
\usepackage{epstopdf}
\usepackage{threeparttable}

\slugcomment{}

\shorttitle{Tidally distorted exoplanets}
\shortauthors{Burton et al.}

\begin{document}

\title{Tidally distorted exoplanets : density corrections for 
short-period hot-Jupiters based solely on observable parameters}

\author{J. R. Burton\altaffilmark{1}, C. A. Watson\altaffilmark{1}, 
A. Fitzsimmons\altaffilmark{1}, D. Pollacco\altaffilmark{2},\\ 
V. Moulds\altaffilmark{1}, S. P. Littlefair\altaffilmark{3},
P. J. Wheatley\altaffilmark{2}}

\email{jburton04@qub.ac.uk}

\altaffiltext{1}{Astrophysics Research Centre, Queen's University Belfast, Belfast, BT7 1NN, UK}
\altaffiltext{2}{Department of Physics and Astronomy, University of Warwick, Coventry, CV4 7AL, UK}
\altaffiltext{3}{Department of Physics and Astronomy, University of Sheffield, Sheffield, S3 7RH, UK}

\begin{abstract}
The close proximity of short period hot-Jupiters to their parent star means they are subject to extreme tidal forces. This has a profound effect on their structure and, as a result, density measurements that assume that the planet is spherical can be incorrect. We have simulated the tidally distorted surface for 34 known short period hot-Jupiters, assuming surfaces of constant gravitational equipotential for the planet, and the resulting densities have been calculated based only on observed parameters of the exoplanet systems. Comparing these results to the density values assuming the planets are spherical shows that there is an appreciable change in the measured density for planets with very short periods (typically less than two days). For one of the shortest-period systems, WASP-19b, we determine a decrease in bulk density of 12\% from the spherical case and, for the majority of systems in this study, this value is in the range of 1-5\%. On the other-hand, we also find cases where the distortion is negligible (relative to the measurement errors on the planetary parameters) even in the cases of some very short period systems, depending on the mass ratio and planetary radius. For high-density gas-planets requiring apparently anomalously large core masses, density corrections due to tidal deformation could become important for the shortest-period systems.
\end{abstract}


\section{Introduction}\label{sec:intro}

Since the first planet outside the solar system was discovered 
orbiting the millisecond pulsar PSR1257+12 (\citealt{wolszcan92}), 
the study of extra-solar planets has rapidly become one of the key 
research areas in astronomy today. At the time of writing, 1782 exoplanets 
have been identified through many different techniques including 
radial velocity measurements, gravitational microlensing, transit surveys 
and many others. As of writing, 1133 are known to be transiting, mostly 
discovered by transit surveys such as SuperWASP (\citealt{pollacco06}), 
CoRoT (\citealt{borde03}), the HATnet project (\citealt{bakos02}) and Kepler 
(\citealt{boricki10}). Transiting systems provide a radius, and hence bulk 
density when coupled with the planetary mass determined from radial 
velocity follow-up, as well as characterising their atmospheres with 
techniques such as transmission spectroscopy (\citealt{charbonneau01},
\citealt{snellen08}) and secondary eclipse photometry (\citealt{gibson10}, 
\citealt{burton12}). The bulk density is an important parameter to determine, 
as this value is used in chemical composition models and atmospheric 
simulations. The bulk density is measured by determining the radius of 
the planet from the transit lightcurve, and assuming a spherical planet 
in order to work out the volume. Since the volume and mass of the planet 
are then both known, a simple bulk density calculation can be applied.

Previous work on modelling exoplanet distortion has been carried out by 
\citet{leconte11}, who demonstrated that tidal and rotational forces result 
in a systematic underestimation of the planetary radius. The tidal distortion 
is particularly relevant for short period planets with low masses and large 
radii, as these systems are subject to a significant amount of gravitational 
interaction with the host star. Since the bulk of the distortion would lie along 
the axis connecting the star and planet, this effect would not be apparent 
during transit observations -- where the planet will appear largely spherical 
due to the observer being presented with a smaller cross-sectional area 
during transit. Failing to account for this distortion would therefore result in a 
systematic overestimation of the measured densities of exoplanets, since the 
volume is calculated based on the planet being spherical. Since a significant 
proportion ($\sim$30\%) of exoplanets so far discovered have orbital separations 
of $\lesssim$0.1A.U. (\citealt{matsumura10}), this could potentially affect the 
measured bulk densities of a number of hot-Jupiter systems. Indeed, work by 
\citet{lammer09} and \citet{li10} show that distortion in WASP-12b is appreciable, 
and can potentially impact on mass-loss rates for short-period systems. Naturally, 
if the tidal forces are weak, the distortion will be negligible, and the planet remains 
close to spherical. 

While \citeauthor{leconte11} are able to calculate the distorted shapes of exoplanets, 
which has important implications for our understanding of the internal dynamics of 
short period systems, their approach is limited in that the models are difficult to reconcile
with observational parameters. For example, the work of \citeauthor{leconte11} requires 
detailed internal structure models, incorporating effects such as rotation and atmospheric 
structure. These parameters are obtained from models derived from measurements of 
Jupiter and Saturn. Since our Solar System's gas giants have markedly different rotation 
periods, orbital separations and incident fluxes to those of the short-period hot-Jupiters, 
accurate determination of such internal parameters for hot-Jupiters is fraught with uncertainty 
and potential error, as well as being beyond the expertise of the average observer. This is 
further demonstrated by the fact that, to the best of our knowledge, the important work by
\citeauthor{leconte11} has never been applied to determine planetary density corrections in 
any observational exoplanet characterisation publication (though it has been used to indicate 
the importance of tidal distortion effects e.g. \citealt{mancini13}, \citealt{kovacs13}). The aim, 
therefore, was to outline a much simpler model to estimate planetary distortion, which would
be easier to implement and would be comparable to the more detailed simulations of 
\citeauthor{leconte11} In order to make the models as applicable to observers as possible, 
the main focus of the modelling was to estimate the distortion using observed parameters 
gleaned from transit photometry and RV measurements.

In this paper, we model the gravitational interaction between the parent star and planet using 
the Roche approximation, and show that tidal distortion can be significant for hot-Jupiters
with orbital periods less than $\sim$2 days. In section 2, we outline the method of constructing 
a distorted hot-Jupiter and measuring the volume in order to work out the change in bulk
density, and use this model in section 3 to estimate the distortion of hot-Saturn-like planets. 
In section 4, we use the parameters of the shortest-period systems, derived from 
follow-up photometry and radial-velocity measurements, to calculate the tidal distortion of 
these exoplanets, and discuss the significance of these findings, comparing the results
with the simulations by \citeauthor{leconte11} 

\section{Method}

In order to model the exoplanets' tidally distorted surface, we have assumed that it lies 
along a surface of constant gravitational potential. Under the Roche approximation 
(\citealt{chandrasekhar69}), the total gravitational potential ($\Phi$) at any point within 
the system is given by;

\begin{equation}
\begin{split}
\Phi = &-\frac{GM_1}{(x^2+y^2+z^2)^{1/2}}-\frac{GM_2}{[(x-a)^2+y^2+z^2]^{1/2}}\\
&-\frac{1}{2}\Omega^2[(x-{\mu}a)^2+y^2],
\label{eqn:Phi}
\end{split}
\end{equation}

\vspace{3mm}

\noindent (e.g. \citealt{pringle85}) where ${M_1}$ is the mass of the primary 
component (in this case the host star) and ${M_2}$ is the mass of the secondary 
(in this case the exoplanet). ${x}$, ${y}$ and ${z}$ are co-ordinates in which the 
surface is plotted with the origin at the centre-of-mass of the primary component, 
where ${x}$ is along the line-of-centre, ${y}$ is  along the orbital plane and ${z}$ 
makes a right-handed set. ${a}$ is the orbital separation between the centres-of-mass 
of the two components. \begin{math}\mu=\frac{M_2}{M_1+M_2}\end{math} and 
\begin{math}\Omega=\frac{2\pi}{P_{orb}}\end{math} (the angular velocity of the system),
where ${P_{orb}}$ is the orbital period.

The Roche approximation has widely been used to model interacting binary systems, 
such as Cataclysmic Variables (CVs). For example, Roche tomography (for a description, 
see e.g. \citealt{rutten96}, \citealt{watson04}) uses the Roche approximation in order to 
reconstruct the surface features of secondaries for a number of CV systems from the 
velocity profiles (\citealt{watson01}, \citealt{watson03}, \citealt{dunford12}). Given the 
extreme tidal forces acting on some short period hot-Jupiters, it was decided to model 
their shapes using the Roche approximation. The main advantage of this approach is 
that it is possible to obtain the distortion solely from the observed system parameters 
as opposed to methods which use parameters derived from simulations, 
such as planetary rotation 
and internal structure. We note here that the Roche approximation has previously been 
used by \cite{budaj11} to estimate the reflection effect for tidally distorted planets, but 
density corrections were not calculated for these systems. 

In order to measure the distortion of the planet using the Roche approximation,
a number of assumptions have been made. First, we have assumed the 
planet is tidally locked with the star, so the orbital period is equal to the rotation 
period. We also assume the orbit has been circularised, so as to keep the distortion
over the orbital period constant. This is a reasonable assumption, as it follows  
predictions of hot-Jupiter formation and follow-up work (e.g. \citealt{nagasawa08}). It is
also a requirement of equation \ref{eqn:Phi}. 
We have also assumed that the mass of the planet is centrally condensed, again, 
required by equation \ref{eqn:Phi}. This was also an assumption made by \citeauthor{leconte11} 
and, for example, simulations by \citet{campo10} predict a centrally condensed WASP-12b 
given an orbital eccentricity of $\sim$0.

\begin{figure}[!h]
\begin{center}
\includegraphics[scale=0.6,trim = 1mm 25mm 1mm 30mm,clip=true]{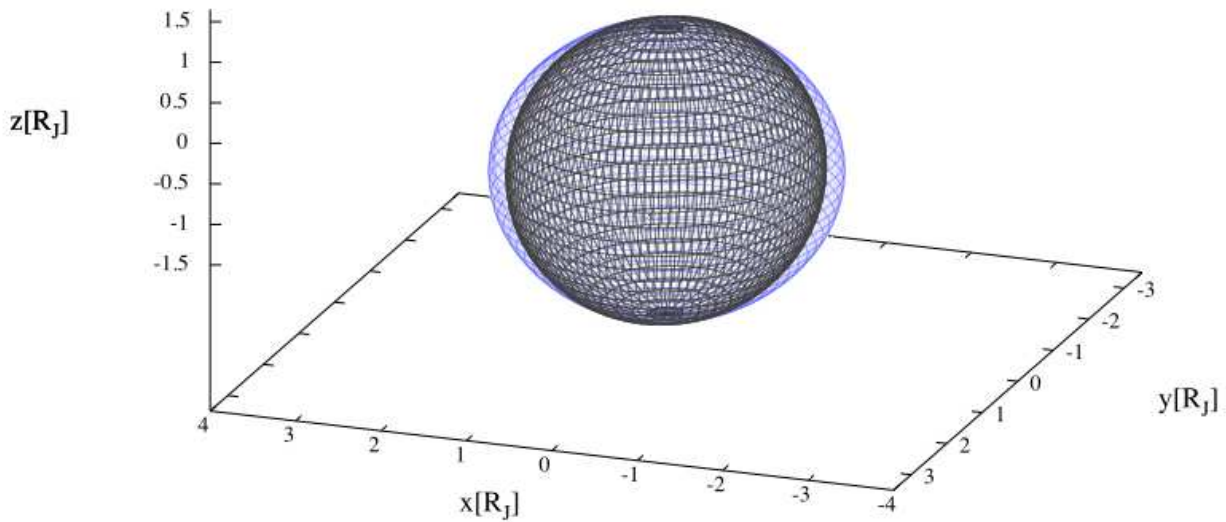}
\caption{The surface of WASP-19b at quadrature for both the spherical case (black line) 
and distorted case (blue, or light grey line), viewed slightly above the orbital plane for 
clarity. Note how the distorted case is more oblate, and the increase in volume is appreciable.}
\label{Roche}
\end{center}
\end{figure}

\begin{figure}[!h]
\begin{center}
\includegraphics[scale=0.6,trim = 1mm 35mm 1mm 1mm, clip=true]{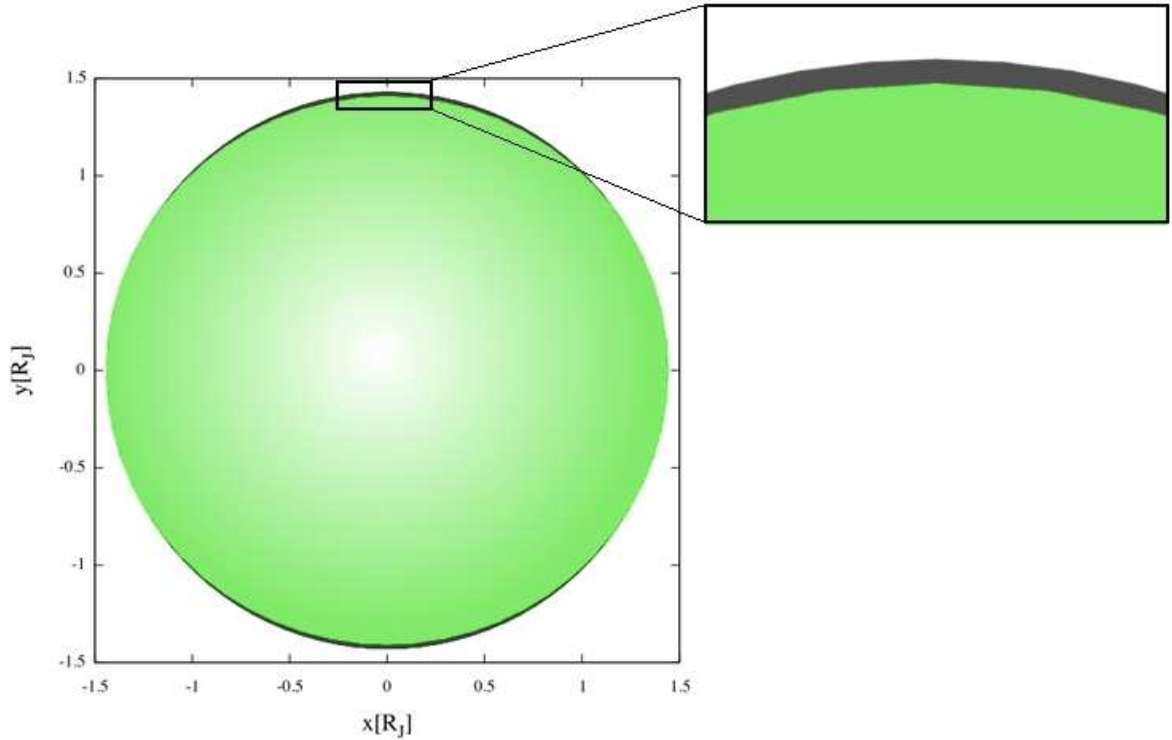}
\caption{The sky projected surface of equal area of WASP-19b at zero phase for both 
the spherical case (black) and distorted case (green/light grey). It can be seen that the 
distorted case is very close to the spherical case, and thus the distortion will be extremely 
difficult to detect during transit.}
\label{zerophase}
\end{center}
\end{figure}

In order to match the model to the observed radius, the value of $\Phi$ in 
equation~\ref{eqn:Phi} was adjusted such that the projected area of the model 
planet during transit matched those given by observations. This method gives 
the quantity ($R_P$/$R_S$)$^2$, the ratio of the planetary occulting area to 
projected star area. Since this is equivalent to the actual parameter derived 
from the transit light curve, this was determined to be the best method in which 
to set the value of $R_P$, as opposed to just setting the value of $\Phi$ to 
return the corresponding polar radius. The three-dimensional output is given in 
Figure \ref{Roche}, along with the spherical case using the parameters of WASP-19, 
a planetary system with one of the shortest known orbital periods (0.79 days). 
Figure \ref{zerophase} shows the sky-projected surface area of WASP-19b for 
both the spherical and distorted case at phase 0. 
As can be seen from the projections, the polar radius at zero phase is almost 
identical. This is due to the bulk of the distortion being projected away from the 
observer, towards the star, making the planet appear spherical in both cases. 

\begin{figure}[!h]
\begin{center}
\includegraphics[angle=270,scale=0.6,trim=1cm 0.8cm 0cm 0cm,clip=true]{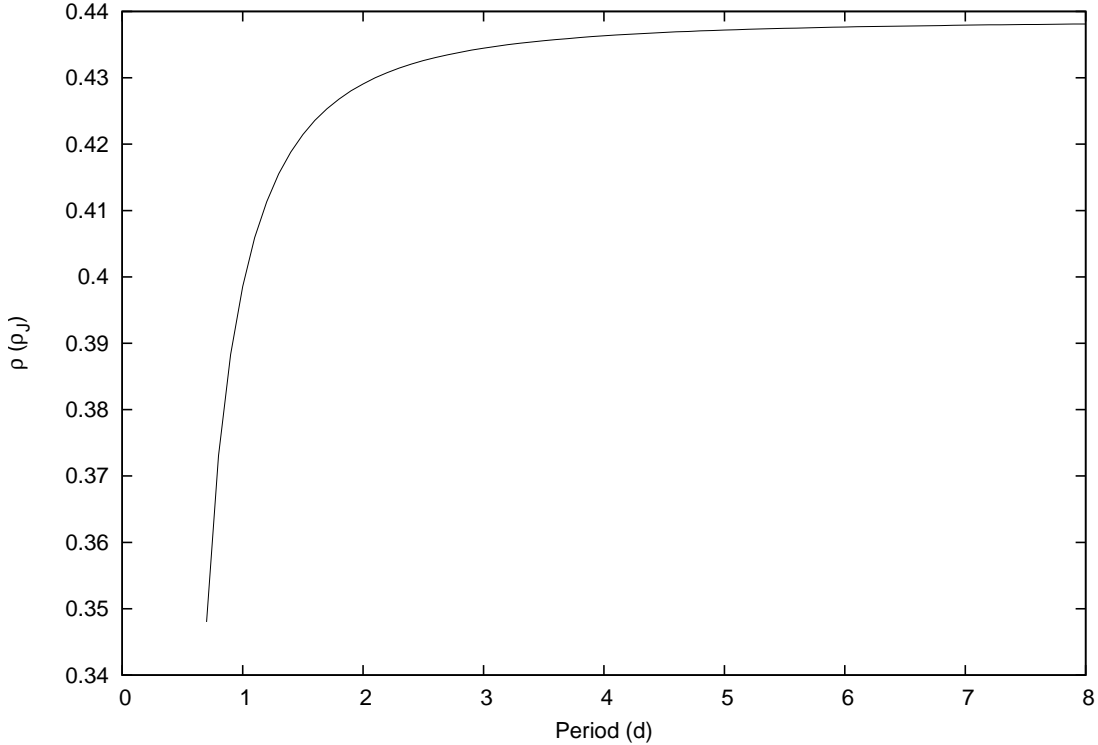}
\caption{Change in bulk density as a function of orbital period using the parameters 
of WASP-19b. After approximately 2 days, the change is negligible when compared 
to the density at the actual period of 0.7888 days.}
\label{track}
\end{center}
\end{figure}

Since the precise co-ordinates of the tiles were known, the volume of 
each `slice' of elements was determined by finding the area bounding 
the surface of the tile with the centre of the slice. The area was then 
multiplied by the depth of each slice, effectively modelling it as a tapered 
wedge. The number of tiles were set sufficiently high to make the gaps 
between sequential tile areas negligible, but also sufficiently low to avoid 
truncation issues when working out the projected area. To check there 
were no jumps in the projected area due to the tile projections, the density 
of the planet was calculated as a function of orbital period. Figure \ref{track} 
is an example of this process using the system parameters of WASP-19b.

This allowed for the internal consistency of the code to be tested by setting 
the orbital period of the system to a value where one would expect the 
distortion to be negligible, and hence the planet to be spherical. At around 
10 days, the difference between the volume given by the spherical case and 
the volume output from the simulations was $\sim$0.001\% for WASP-19b. 
Once the accuracy and consistency of the code had been thoroughly
established and tile truncation and placement errors had been minimised, 
the parameters of thirty-four short-period hot-Jupiters were 
obtained in order to model the tidal distortion for each of these systems.
We discuss the impact of this
in section \ref{results}.

\section{Saturn-like planets}

As a final demonstration, a number of cases of exoplanets with parameters 
similar to Saturn have been modelled with the orbital period set to one day, 
the results of which are summarised in Table \ref{tab:Simulations of Saturn-like planets}. 
This was carried out in order to investigate the region of planetary
parameters, primarily star-to-planet mass ratio, which are most susceptible
to a severe (i.e. $>$20\%) level of distortion.
A 1$M_J$ model and 0.3 $M_J$ (Saturn-like) model have been added as a 
comparison. We have also run the simulations over a range of orbital periods. 
The results from these simulations indicate that for Saturn-mass planets, the 
density correction is significantly important for orbital periods of less than 
$\sim$1.5 days. For a hot-Saturn with an orbital period of 0.6 days, the density 
correction approaches 25\%, sufficient to alter the position of the exoplanet on 
the mass-radius relation (see Figure \ref{fig:massrad}). Whilst there has yet to 
be a candidate discovered matching these orbital parameters, many systems 
are being identified with densities matching that of Saturn with short orbital periods 
(e.g \citealt{fortney11}, \citealt{ojeda13}, \citealt{gandolfi13}). It can be seen that a 
density correction of $\sim$5-10\% is being approached for systems with more 
massive host stars, for orbital periods of 1 day. Whist systems matching these exact
parameters have yet to be discovered, hot-Jupiters with orbital periods of less than one
day are known to exist (e.g. WASP-18b, WASP-19b, CoRoT-7b, Kepler-10b,
Kepler-70b). The parameters we have used for these simulations provide an initial, if
somewhat speculative, investigation into planets which could potentially show
extreme levels of distortion.
In the case of GJ163b -- a 
non-transiting exoplanet -- we have simply used the mass ratio of the system with 
a Saturn-like radius, as the radius of the planet cannot be constrained from radial 
velocity measurements. 

An especially interesting case is that of HD149026b. 
Since its discovery in 2005, the incredibly high density (1.17gcm$^{-3}$) has 
resulted in the prediction of an anomalously high core mass of 
70-85$M_{\oplus}$ (\citealt{sato05}). Previous predictions of planetary formation
using the core accretion model indicate a maximum core mass of 30$M_{\oplus}$ 
before runaway gas accretion occurs (\citealt{ikoma06}). Several explanations for 
the massive core of HD149026b have been proposed, including heavy element 
rain onto the core, accretion of planetesimals and collisions between planetary
embryos (\citealt{sato05}; \citealt{fortney06}). While the gravitational distortion of 
this planet with its current orbital period of 2.8 days is insufficient to lower its density 
in order to reduce the core mass to levels that agree with accretion theory,
gravitational distortion could allow this for shorter period systems, should they be 
observed. For example, a Saturn-mass planet in a 1 day orbit around a 1.1$M_{\odot}$ 
star would yield a radius of 0.7$R_{J}$ if sphericity was assumed, but would actually 
have a mean radius closer to 0.8$R_{J}$ due to gravitational distortion. Using the 
evolutionary models of \cite{bodenheimer03} in addition to the core mass values 
calculated by \cite{sato05} with our new density value would allow a more accurate 
core mass to be estimated for such systems. For the case above, assuming a spherical 
planet (i.e. a radius of 0.7$R_{J}$), the core mass would be calculated to be 85$M_{\oplus}$ 
assuming a core density of 5.5gcm$^{-3}$, and a core mass of $\textless$74$M_{\oplus}$ 
assuming a core density of 10.5gcm$^{-3}$ (the core density of Saturn). The value of our 
density correction, however, would result in a corrected estimate of the core mass of 
$\textless$74$M_{\oplus}$ assuming a core density of 5.5gcm$^{-3}$, and a core mass 
of $\textless$60$M_{\oplus}$ assuming a core density of 10.5gcm$^{-3}$ for the distorted 
case. This reduction in the core mass of a planet analogous to HD149026b on a sufficiently 
short-period indicates that hot-Saturns discovered to have a higher than expected core mass 
can be partially explained by the tidal distortion effect. Additional mechanisms, such as 
accretion of planetesimals, can then be invoked to explain this much more easily with a 
reduced core mass. 

\vspace{5mm}
\begin{table}[h!]
\begin{center}
\footnotesize
\caption[Simulations of Saturn-like planets]{Output from simulations using the mass-ratios of 
three exoplanet systems, HD149026, WASP-29 and GJ163, using a 1 day orbital period. 
Also included
as a comparison are the cases for a Jupiter-mass planet and a Saturn-mass planet orbiting
a 1$M_{\odot}$ star with a period of 1 day. The density values have been indicated for both 
the spherical (6) and distorted cases (7).}
\label{tab:Simulations of Saturn-like planets}
\begin{threeparttable}
\begin{tabular}{lccccccc}
\hline
System & $M_S$ ($\mathrm{M}_{\odot}$) & $M_P$ ($\mathrm{M}_{\rm J}$) & $R_P$ ($\mathrm{R}_{\rm J}$) & Ref. & ${\rho}_{sph}$ & ${\rho}_{1d}$ & ${\Delta}{\rho}$ \\
(1) & (2) & (3) & (4) & (5) & (6) & (7) & (8) \\
\hline
Model hot-Jupiter & 1.0 & 1.0 & 1.0 & - & 0.238 & 0.229 & 3.0\% \\
Model hot-Saturn & 1.0 & 0.3 & 0.8 & - & 0.186 & 0.172 & 7.5\% \\
HD149026b & 1.3 & 0.356 & 0.718 & [1] & 0.229 & 0.210 & 8.25\%  \\
WASP-29b & 0.825 & 0.244 & 0.792 & [2] & 0.117 & 0.108 & 7.85\% \\
GJ163b & 0.4 & 0.354 & 0.8\tnote{$^*$} & - & 0.156 & 0.163 & 4.47\% \\
\hline
\end{tabular}
\begin{tablenotes}
\item[] [1]  \cite{southworth10} [2] \cite{hellier10} \\
\item[$^*$]Since this candidate is not a transiting planet, only the mass has been 
constrained from radial velocity measurements. For this planet, a radius comparable 
to that of Saturn has been assumed.
\end{tablenotes}
\end{threeparttable}
\end{center}
\end{table}

\section{Results and discussion}\label{results}

\begin{figure}[h!]
\begin{center}
\includegraphics[scale=1,trim=0mm 0mm 0mm 0mm,clip=true]{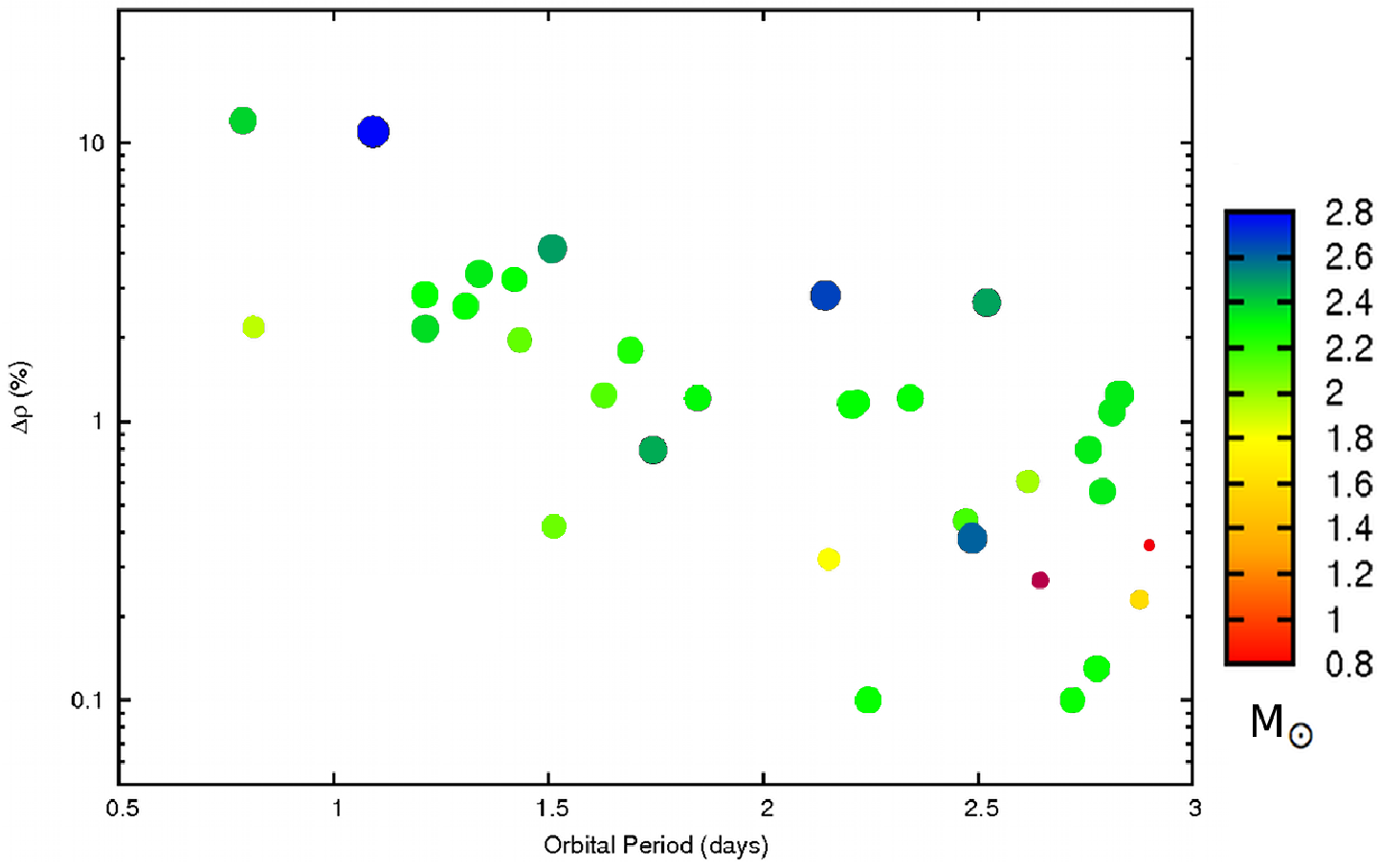}
\rule{15em}{0.25pt}
\caption[Decrease in bulk density as a function of orbital period for the simulations]{The 
decrease in bulk density on a logarithmic scale as a function of orbital period for the selection
of short-period hot-Jupiters from Table \ref{tab:Density corrections for 34 short-period hot-Jupiter systems}. 
As expected from the 
Roche approximation (equation \ref{eqn:Phi}), as the orbital period increases, the density 
correction decreases. The respective size of the circles represent the radius of the 
planet, and the colour gradient represents the mass of the host star.}
\label{fig:no}
\end{center}
\end{figure}

The results for thirty-four short-period hot-Jupiters are shown in Table 
\ref{tab:Density corrections for 34 short-period hot-Jupiter systems}. 
The most significant distortion of the hot-Jupiters modelled occur for the 
shortest period systems, with WASP-19b having a density which is 12\% 
less than the spherical case. Since the size of the Roche lobe is dependent 
on the period (and thus the separation distance of the components) in addition
to the mass ratio of the two components, this 
follows expectations. However, WASP-43b also has an extremely short 
period, and one therefore might expect a distortion comparable to that of 
WASP-19b or WASP-12b ($\sim$11\%). Whilst the period is a key parameter 
in the size of the critical Roche lobe, in the case of WASP-43b the density is 
extremely high (over twice that of Jupiter). This means that for the mass ratio 
of this system, combined with the relatively small radius of the planet, the 
distortion due to tidal interactions for this system is negligible, at least in 
comparison with systems with a similar orbital period. Essentially, for planets 
with a large mass ($>$1.5$M_J$), but with a radius comparable to that of 
Jupiter, the distortion is negligible for a given stellar mass. 
Again, this follows expectations, since 
the critical Roche lobe is much bigger and, as a consequence, the planet 
has very little distortion as it lies well within its Roche lobe. This can be seen 
for planets such as Corot-14b and WASP-14b. Also, for planets, with an orbital 
period of greater than $\sim$2 days, the distortion becomes negligible as is also 
evident in Figure \ref{track}.  The other main variable which causes a similar change
in the size of the Roche lobe is the mass ratio of the two components. For a 
more massive host star, the size of the critical Roche lobe decreases. 
Conversely, a more massive planet will allow for the
size of the Roche lobe to increase, however in the case of the extreme mass ratio
between a host star and planet, this effect is much less significant than for example
in CVs, where the mass of the components is comparable. For the planets we have
investigated, the mass ratio of the systems is relatively uniform, as indicated in Figure
\ref{fig:no} and Table \ref{tab:Density corrections for 34 short-period hot-Jupiter systems}.
However, given the extreme mass ratio between hot-Jupiter and host star (of the order
$\sim$10$\times$$^{-3}$), this parameter value is a very important determining value
as to the distortion of the exoplanet.
For the majority of systems we 
have modelled, the change in bulk density from the spherical case is in the range 
1-5\%, meaning that even for systems where the parameters are not particularly 
conducive to a major distortion, this effect can still be significant. It should be noted 
that typical measurement uncertainties on the masses and radii of exoplanets (and their
host stars) lead 
to errors on the derived bulk densities of the order of $\sim$5\%. Thus, for the majority 
of the systems in Table \ref{tab:Density corrections for 34 short-period hot-Jupiter 
systems} the correction required is not significant -- and only the most heavily distorted 
planets present a significant level of bulk density correction. In the specific case of 
WASP-19b, for example, the measurement error on the currently published bulk 
density is 6.1\% (\citealt{hellier11a}) -- significantly less than the correction needed 
to account for tidal distortion (12.0\%). In addition, it should be noted that the distortion 
effect is systematic in that it can only decrease the density. In the future, as measurement
errors on the radius and mass of the systems are refined, the bulk density error value/distortion 
will become more important.
Figure \ref{fig:no} shows the decrease in bulk density as a function of orbital 
period for these planets from Table \ref{tab:Density corrections for 34 short-period hot-Jupiter systems}. The radius of the hot-Jupiter is represented by the 
size of the points on the graph. As expected, the shortest period hot-Jupiters 
are the most susceptible to tidal deformation, leading to overestimates of the 
planetary density of $>$10\% in some cases. 

The results from our simplified Roche models agree with the more complex approach of 
\citeauthor{leconte11}. The two cases of particular interest being the two most distorted 
exoplanets we modelled; WASP-12b and WASP-19b. \citeauthor{leconte11} return a 
radius increase in the direction of the distortion of 3\% for WASP-12b and 2.72\% for 
WASP-19b, in comparison with our volume calculations which return an increase 
of 3.21\% and 3.56\%, respectively. These figures are likely to change as the error
bars on the observed parameters are refined, and the difference between the radius values
calculated from simulations (as used by \citeauthor{leconte11}) and the simpler Roche
method should be compared when new volume models are generated.
It is also mentioned by \citeauthor{leconte11} that 
WASP-12b will have a mean density decrease of $\sim$9\% based on their calculations, 
again, in close agreement with our value of 11\%. The models described in this paper 
provide observers with a fast and simple estimation of the distortion, without the need 
to invoke estimations of uncertain internal parameters such as rotation rates, or requiring 
the user to calculate complex internal structure profiles. Indeed, recent observations by 
\cite{sing13} have indicated that WASP-12b has a density $\sim$20\% lower than 
previously estimated. By comparing the results from our model with observations such 
as these, we will be able to improve the accuracy of the density correction estimation by 
adding effects such as mass loss during subsequent iterations.

We must also consider the limitations of using this technique to model the distortion. 
As previously mentioned, equation \ref{eqn:Phi} assumes orbital circularisation, and 
while this is a reasonable assumption, two of the planets considered in this work have 
significant eccentricities. HAT-P-23b and GJ436b both have $e$$\sim$0.1, meaning the 
method used for these systems is not strictly applicable, and the density changes for 
these planets will differ from the stated amounts. For GJ436, this is not much of an issue, 
as the spherical density of this planet is already high ($>$1.5$\rho_J$), and the distortion 
given its 2.64-day period will not be significant. However, for systems with periods 
comparable to that of HAT-P-23b (1.2 days) and with appreciable eccentricities, the 
method we have used will produce a density change which is likely to be incorrect.
The system components are also assumed to be centrally condensed. Again, while this 
is a reasonable assumption, work carried out by \citet{sirotkin09} indicates that the internal 
structure of low mass CV donors deviate from this assumption, resulting in slightly different 
Roche-volumes on the level of a few percent. While the models we have constructed do not 
fill the critical Roche lobe, this is nevertheless an important effect, as planets with a sufficiently
 high density only have a volume increase on the level of 0.5-1\%.

As stated by \citet{li10}, the light curve from a tidally-distorted hot-Jupiter
would be different from the spherical case. Simulations indicate that at the quarter
and three-quarters phase, the flux would be $\sim$10\% greater due to
the effective area being greater. Sufficient observations of day-night variations
in the infra-red could, in theory, be able to detect this. However, since the change in intensity
is almost beyond detection limits for ground-based platforms, space-based observations
would be the ideal approach to detect the distortion (e.g. Spitzer, Hubble).
In addition, oblateness of a transiting exoplanet would reveal itself through slight
anomalies in ingress and egress. \citet{carter09} used Spitzer photometry
to place oblateness constraints on the hot-Jupiter HD189733, concluding it is
less oblate than Saturn. This is to be expected, however, as the spin rate for
hot-Jupiters are expected to have been slowed due to tidal friction into synchronous orbit.
It has also been shown that this effect is extremely difficult to measure, even for hot-Jupiters
with periods of less than two days (\citealt{carter10}). From Figure \ref{Roche}, when viewed 
from the quarter-phase position, the distortion of the planet does indeed appear appreciable. 
Figure \ref{zerophase}, however, shows the distortion at 90$^{\circ}$ inclination and zero phase. 
From this angle, the two cases are remarkably similar, indicating that even for planets with an 
appreciable distortion the oblateness is not particularly significant, in agreement with \citet{carter10}.

\begin{figure}[h!]
\begin{center}
\includegraphics[scale=0.6,trim=5mm 0mm 0mm 0mm,clip=true]{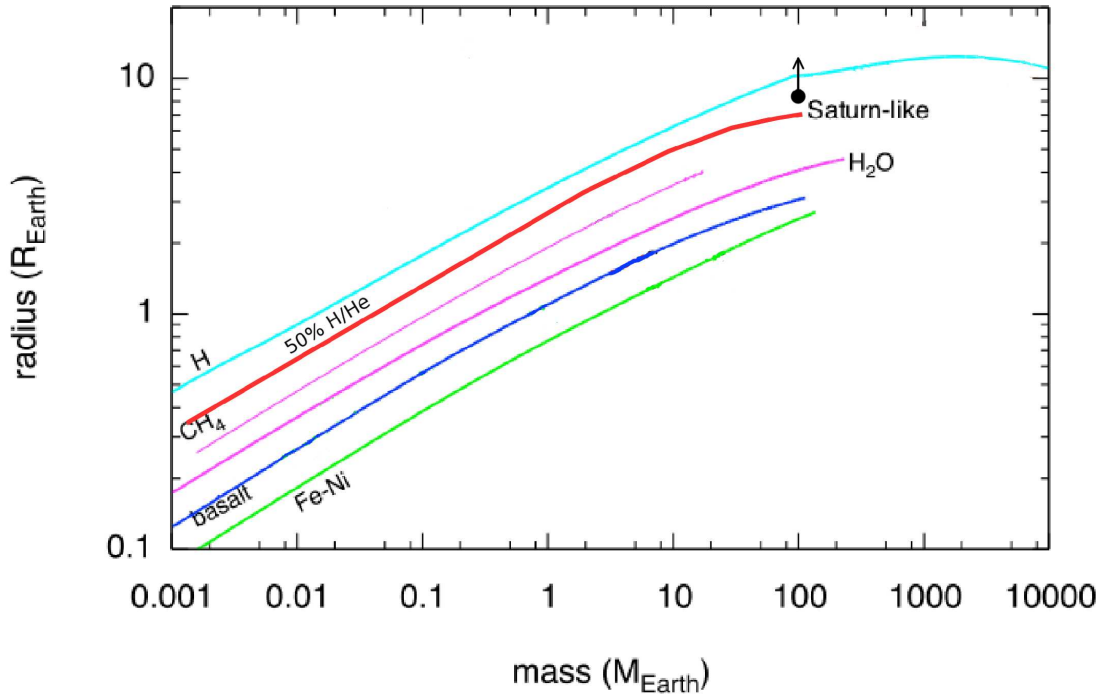}
\rule{15em}{0.25pt}
\caption[The effect of distortion on models of bulk composition]{Mass-radius relations for a 
number of different compositions from \cite{swift12}. The 50\% H/He line (red curve) has been 
added from \cite{sasselov08}.  The arrow represents the magnitude by which the tidal 
distortion can alter the radius from the spherical case.}
\label{fig:massrad}
\end{center}
\end{figure}

Given the likely over-estimation of the density of short-period hot-Jupiters as currently 
reported, there may also be an impact on current atmospheric simulations, as the species 
of molecule composing the upper atmosphere is dependent on the overall density.
For example, \citet{adams08} and \citet{swift12} show that for a given mass-radius relation, 
the exoplanet's bulk composition can be found. However, the bulk composition is insensitive 
to a change in mass for planets greater than $\sim$100 Earth masses ($\sim$0.3$M_J$) but 
is still sensitive to a change in effective radius. Mass-density plots, such as those provided in
\cite{szabo11} for transiting exoplanets will also require an adjustment factor for the lower-mass 
hot-Jupiters. A change of 10\% of the volume of a planet may allow for the planet to move from a
bulk composition of H/He to one of H alone. Simulations indicate that a Saturn-mass planet with 
a radius of $\sim$1$R_J$ would have a volume 10\% greater than an assumed spherical planet 
on a 1.5-day-period, meaning that for hot-Saturns, a shift in the bulk composition is possible.
Figure \ref{fig:massrad} shows an example of the mass-radius parameter space from \cite{swift12}
highlighting the particular area of interest which would result in a potential impact on composition 
simulations.

\section{Conclusions}

The tidal forces short-period exoplanets are subject to due to gravitational interactions 
with the host star can result in an appreciable decrease in the bulk density once this 
effect is accounted for. Due to the bulk of the distortion being along the axis connecting 
the centres-of-mass, this distortion is not apparent during transit, and hence when the 
density of the planet is being measured, this effect is not accounted for. For the most 
distorted systems, the corrected bulk densities could cause the assumed bulk 
composition to change, meaning atmospheric simulations must take this into account 
when dealing with short-period systems, as well as for planets with a low mass/radius 
ratio. 

We have demonstrated that for short-period systems, the decrease in bulk density 
from the spherical case is of the order of 10\% for systems where the parameters 
allow for the most distortion. It has also been shown that for particularly dense 
planets, this effect is negligible, even for systems with periods less than one day. 
As further exoplanets are discovered, the mass-ratio of these systems will also provide
a further census of planets with more extreme mass ratios where the gravitational distortion
will be enhanced.
The method we have used matches previous models based on much more complex 
internal modelling of the planet as a whole, and provides a simple method for 
observers to estimate the distortion limits. As more hot-Jupiters with even more 
extreme environments are being discovered, it is imperative that any correction in 
the parameters obtained for these systems are applied in order to obtain the most 
accurate possible atmospheric models. In addition, modelling hot-Saturn
planets can also provide important information for internal modelling simulations, 
as in the case of a short-period planet with parameters similar to HD149026b.

\section*{\sc Acknowledgements}

JB is funded by the Northern Ireland Department of Employment and learning, 
and in addition would like to thank C Walsh, M Fraser and M McCrum for useful 
discussion.

\begin{table}[h!]
\small
\begin{center}
\caption[Density corrections for 34 short-period hot-Jupiter systems]{Derived parameters for 
short-period hot-Jupiter exoplanets, with modified density values taking into account distortion
due to tidal forces. Columns 1-6 and 8 are the parameters given in the corresponding reference (7). 
Column 9 is the adjusted density for the planet taking into account tidal distortion. Column 10 is
the percentage difference between the two densities.}
\label{tab:Density corrections for 34 short-period hot-Jupiter systems}
\begin{threeparttable}
\begin{tabular}{lccccccccr}
\hline
System & P(d) & $M_S$ ($\mathrm{M}_{\odot}$) & $M_P$ ($\mathrm{M}_{\rm J}$) & $R_P$ ($\mathrm{R}_{\rm J}$) &  $i$ ($^{\circ}$) & Ref. & ${\rho}_{sph}$ & ${\rho}_{dis}$ & ${\Delta}{\rho}$\\
(1) & (2) & (3) & (4) & (5) & (6) & (7) & (8) & (9) & (10)\\

\hline
WASP-19 & 0.7888 & 0.97 & 1.168 & 1.386 & 79.4 & [1] & 0.438 & 0.392  & 12.00\% \\
WASP-43 & 0.8134 & 0.58 & 1.78 & 0.93 & 82.6 & [2] & 2.210 & 2.160 & 2.18\% \\
WASP-12 & 1.0914 & 1.35 & 1.41 & 1.79 & 83.1 & [3] & 0.240 & 0.222 & 10.95\% \\
OGLE-TR-56 & 1.2119 & 1.17 & 1.29 & 1.30 & 78.8 & [4] & 0.587 & 0.571 & 2.85\% \\
HAT-P-23\tnote{$^*$} & 1.2129 & 1.13 & 2.090 & 1.368 & 85.1 & [5] & 0.816 & 0.799 & 2.16\% \\
TrES-3 & 1.3061 & 0.924 & 1.92 & 1.295 & 82.15 & [6] & 0.884 & 0.872 & 2.59\% \\
Wasp-4 & 1.3382 & 0.93 & 1.250 & 1.34 & 89.47 & [7] & 0.520 & 0.502 & 3.39\% \\
Qatar-1 & 1.4200 & 0.85 & 1.090 & 1.165 & 83.47 & [8] & 0.689 & 0.676 & 3.23\% \\
OGLE-TR-113 & 1.4325 & 0.78 & 1.24 & 1.11 & 87.7 & [9] & 0.907 & 0.889 & 1.96\% \\
Corot-1 & 1.5090 & 0.95 & 1.03 & 1.49 & 85.1 & [10] & 0.311 & 0.299 & 4.16\% \\
Corot-14 & 1.5121 & 1.13 & 7.6 & 1.09 & 79.6 & [11] & 5.869 & 5.844 &  0.42\% \\
WASP-5 & 1.6284 & 1.00 & 1.555 & 1.14 & 86.1 & [7] & 1.050 & 1.037 & 1.24\% \\
OGLE-TR-132 & 1.6899 & 1.297 & 1.17 & 1.25 & 83.3 & [12] & 0.599 & 0.588 & 1.80\% \\
Corot-2 & 1.7430 & 0.97 & 3.31 & 1.466 & 87.84 & [13] & 1.050 & 1.042 & 0.79\% \\
WASP-3 & 1.8468 & 1.24 & 1.76 & 1.31 & 84.4 & [14] & 0.783 & 0.774 & 1.21\% \\
WASP-48 & 2.1436 & 1.09 & 0.98 & 1.67 & 80.09 & [15] & 0.21 & 0.205 & 2.83\% \\
WASP-2 & 2.1522 & 0.803 & 0.847 & 0.807 & 84.81 & [12] & 1.612 & 1.606 & 0.32\% \\
HAT-P-7 & 2.2047 & 1.47 & 1.800 & 1.42 & 84.1 & [16] & 0.629 & 0.621 & 1.15\% \\
HD189733 & 2.2186 & 0.840 & 1.15 & 1.15 & 85.78 & [12] & 0.756 & 0.747 & 1.17\% \\
WASP-14 & 2.2438 & 1.211 & 7.341 & 1.28 & 84.32 & [17] & 3.500 & 3.490 & 0.10\% \\
WASP-24 & 2.3412 & 1.184 & 1.071 & 1.30 & 83.64 & [18] & 0.487 & 0.481 & 1.21\% \\
TrES-2 & 2.4706 & 0.98 & 1.198 & 1.169 & 84.15 & [19] & 0.750 & 0.747 & 0.44\% \\
OGLE2-TR-L9 & 2.4855 & 1.52 & 4.5 & 1.61 & 79.8 & [20] & 1.078 & 1.074 & 0.38\% \\
WASP-1 & 2.5200 & 1.243 & 0.860 & 1.48 & 88.0 & [12] & 0.265 & 0.258 & 2.67\% \\
XO-2 & 2.6159 & 0.98 & 0.57 & 0.980 & 88.9 & [21] & 0.606 & 0.602 & 0.61\% \\
GJ 436\tnote{$^*$} & 2.6439 & 0.459 & 0.0737 & 0.365 & 86.43 & [12] & 1.516 & 1.512 & 0.27\% \\
WASP-32 & 2.7187 & 1.10 & 3.60 & 1.18 & 85.3 & [22] & 2.191 & 2.189 & 0.10\% \\
WASP-26 & 2.7566 & 1.12 & 1.02 & 1.32 & 82.5 & [23] & 0.443 & 0.440 & 0.79\% \\
HAT-P-16 & 2.7760 & 1.218 & 4.193 & 1.289 & 86.6 & [24] & 1.958 & 1.955 & 0.13\% \\
HAT-P-5 & 2.7885 & 1.16 & 1.06 & 1.26 & 86.75 & [25] & 0.530 & 0.527 & 0.56\% \\
HAT-P-30 & 2.811 & 1.242 & 0.711 & 1.34 & 83.6 & [26] & 0.295 & 0.292 & 1.08\% \\
Corot-12 & 2.8280 & 1.078 & 0.917 & 1.44 & 85.48 & [27] & 0.309 & 0.303 & 1.25\%  \\
HD149026 & 2.8759 & 1.271 & 0.356 & 0.610 & 88.0 & [12] & 1.570 & 1.560 & 0.23\% \\
HAT-P-3 & 2.8997 & 0.936 & 0.599 & 0.889 & 87.24 & [28] & 0.853 & 0.849 & 0.36\% \\

\hline
\end{tabular}
\begin{tablenotes}
\footnotesize
\item[$^*$]These planets have an appreciable eccentricity, with HAT-P-23 having $e$=0.106, and GJ436
having $e$=0.15, and therefore do not strictly satisfy the condition of orbital circularisation assumed by 
equation \ref{eqn:Phi}.\\
\end{tablenotes}
\end{threeparttable}
\end{center}
\end{table}

\clearpage

\noindent [1]  \citet{hellier11a} [2] \citet{hellier11b} [3] \citet{hebb09} [4] \citet{pont07} [5] \citet{bakos11} [6] \citet{odonovan07} [7] \citet{triaud10} [8] \citet{alsubai11} [9] \citet{gillon06} [10] \citet{barge08} [11] \citet{tingley11} [12] \citet{southworth10} [13] \citet{alonso08} [14] \citet{pollacco08} [15] \citet{enoch11}
[16] \citet{lammer09} [17] \citet{joshi09} [18] \citet{street10} [19] \citet{christiansen11} [20] \citet{snellen09b} [21] \citet{burke07} [22] \citet{maxted10} [23] \citet{smalley10} [24] \citet{buchhave10} [25] \citet{bakos07} [26] \citet{johnson11} [27]  \citet{gillon10} [28] \citet{torres07} \\
\end{document}